\documentclass[twocolumn,superscriptaddress,floatfix]{revtex4}
\usepackage{xr}
\usepackage{chapterbib}
\usepackage[dvips]{graphicx}
\usepackage{latexsym}
\usepackage{amsmath}
\usepackage{amsfonts}
\usepackage{amssymb}
\usepackage{bm}
\usepackage{color}
\usepackage{txfonts}
\usepackage{mhchem}
\begin{document}
	\newcommand{\fig}[2]{\includegraphics[width=#1]{#2}}
	\newcommand{\la}{{\langle}}
	\newcommand{\ra}{{\rangle}}
	\newcommand{\dg}{{\dagger}}
	\newcommand{\upa}{{\uparrow}}
	\newcommand{\dna}{{\downarrow}}
	\newcommand{\ab}{{\alpha\beta}}
	\newcommand{\ias}{{i\alpha\sigma}}
	\newcommand{\ibs}{{i\beta\sigma}}
	\newcommand{\hH}{\hat{H}}
	\newcommand{\hn}{\hat{n}}
	\newcommand{\hc}{{\hat{\chi}}}
	\newcommand{\hU}{{\hat{U}}}
	\newcommand{\hV}{{\hat{V}}}
	\newcommand{\br}{{\bf r}}
	\newcommand{\bk}{{{\bf k}}}
	\newcommand{\bq}{{{\bf q}}}
	\def\gsim{~\rlap{$>$}{\lower 1.0ex\hbox{$\sim$}}}
	\setlength{\unitlength}{1mm}
	\newcommand{{\vhf}}{$\chi^\text{v}_f$}
	\newcommand{{\vhd}}{$\chi^\text{v}_d$}
	\newcommand{{\vpd}}{$\Delta^\text{v}_d$}
	\newcommand{{\ved}}{$\epsilon^\text{v}_d$}
	\newcommand{{\vved}}{$\varepsilon^\text{v}_d$}
	\newcommand{{\tr}}{{\rm tr}}
	\newcommand{\pprl}{Phys. Rev. Lett. \ }
	\newcommand{\pprb}{Phys. Rev. {B}}

\title {Gapless excitations inside the fully gapped kagome superconductors AV$_3$Sb$_5$}
\author{Yuhao Gu}
\thanks{These two authors contributed equally}
\affiliation{Beijing National Laboratory for Condensed Matter Physics and Institute of Physics,
	Chinese Academy of Sciences, Beijing 100190, China}

\author{Yi Zhang}
\thanks{These two authors contributed equally}
\affiliation{Kavli Institute of Theoretical Sciences, University of Chinese Academy of Sciences,
	Beijing, 100190, China}

\author{Xilin Feng}
\affiliation{Beijing National Laboratory for Condensed Matter Physics and Institute of Physics,
	Chinese Academy of Sciences, Beijing 100190, China}
\affiliation{School of Physical Sciences, University of Chinese Academy of Sciences, Beijing 100190, China}

\author{Kun Jiang}
\email{jiangkun@iphy.ac.cn}
\affiliation{Beijing National Laboratory for Condensed Matter Physics and Institute of Physics,
	Chinese Academy of Sciences, Beijing 100190, China}

\author{Jiangping Hu}
\email{jphu@iphy.ac.cn}
\affiliation{Beijing National Laboratory for Condensed Matter Physics and Institute of Physics,
	Chinese Academy of Sciences, Beijing 100190, China}
\affiliation{Kavli Institute of Theoretical Sciences, University of Chinese Academy of Sciences,
	Beijing, 100190, China}
\date{\today}

\begin{abstract}
The superconducting gap structures in the transition-metal-based kagome metal AV$_3$Sb$_5$ (A=K,Rb,Cs), the first family of quasi-two-dimensional kagome  superconductors,  remain elusive as there is  strong experimental evidence for both nodal and nodaless gap structures.   Here we show  that the dichotomy can be resolved  because of the coexistence of   time-reversal symmetry breaking   with a conventional fully gapped  superconductivity.  The symmetry protects  the  edge states which arise on the domains of the lattice symmetry breaking order to remain gapless in proximity to a conventional pairing. We demonstrate this result in a  four-band tight-binding model using the V $d_{X^2-Y^2}$-like  and the in-plane Sb $p_z$-like Wannier functions that can faithfully capture the main feature  of the materials near the  Fermi level.  \end{abstract}

\maketitle

Because of their unique lattice structure,  kagome materials have become an important platform for studying the interplay between electron correlation, topology and geometry frustration. Recently, the family of  AV$_3$Sb$_5$ (A=K,Rb,Cs) materials has been found to be the first quasi-two-dimensional kagome superconductors (SC)\cite{ortiz19,ortiz20,ortiz21,lei,lishiyan,hqyuan,jgcheng,zywang,gaohj}.  These materials display many very intriguing phenomena. For example, an unconventional charge density wave (CDW) order has been found in the non-magnetic AV$_3$Sb$_5$  \cite{ortiz19,ortiz20,graf, topocdw} , which is also concurrent with the anomalous Hall effect (AHE) \cite{xhchen}. 
Muon spin spectroscopy ($\mu$SR) measurements \cite{topocdw,musr1,yuli,musr_sc} have revealed solid evidence for time-reversal symmetry breaking (TRSB). In order to explain the TRSB, many  theories  have been proposed \cite{feng,ronny,nandkishore,balents,feng2}. In particular, the chiral flux phase (CFP) \cite{feng}, which carries unique nontrivial  topological properties,  can naturally explain the TSRB and AHE.

For the superconducting properties of AV$_3$Sb$_5$, there are many controversial experimental results. On one hand, the superconductivity appears to be quite conventional. A Hebel-Slichter coherence peak appears just below the SC $T_c$ from the spin-lattice relaxation measurement from the $^{121/123}$Sb nuclear quadrupole resonance (NQR) \cite{zhengli}.  The coherence peak is widely known as a hallmark for a conventional  $s$-wave SC \cite{hebel1,hebel2}, which is also consistent with the decreasing Knight shift after the SC transition \cite{zhengli}. The magnetic penetration depth measurements also suggest a full gapped superconducting state for CsV$_3$Sb$_5$ \cite{yuanhq,musr_sc}.
There is also no magnetic resonance peak, which normally appears in superconductors with strong electron-electron correlation , such as cuprates and iron-based superconductors \cite{scalapino}. The weak electron-electron correlation in these materials is consistent with angle-resolved photoemission spectroscopy (ARPES) measurements \cite{zhouxj,ortiz20,miaoh,zhangy,sato,hejf,wangsc,shim} and the first-principle calculations \cite{wangyl,luzy}.  In addition, the SC is very sensitive to magnetic impurities but without any resonance peaks to non-magnetic impurities \cite{fengdl}.   The $\mu$SR measurements also fail to detect any additional TRSB signals below $T_c$, which indicates a time-reversal persevered SC order parameter \cite{musr_sc}. On the other hand,
 the  thermoelectric transport measurements show a  finite residual thermal conductivity at $T \rightarrow 0$  in CsV$_3$Sb$_5$, which indicates a nodal SC feature \cite{lishiyan,xxwu}. Scanning tunneling microscopy (STM) observes the V-shape density of states (DOS), which is a typical feature for gapless SC \cite{zywang,gaohj,fengdl}.
 As  the existence of a gapless excitation normally indicates an unconventional superconductor,  it is fundamentally important to find out a reconciliation of these experimental results.   
 
\begin{figure*}
	\begin{center}
		\fig{7.0in}{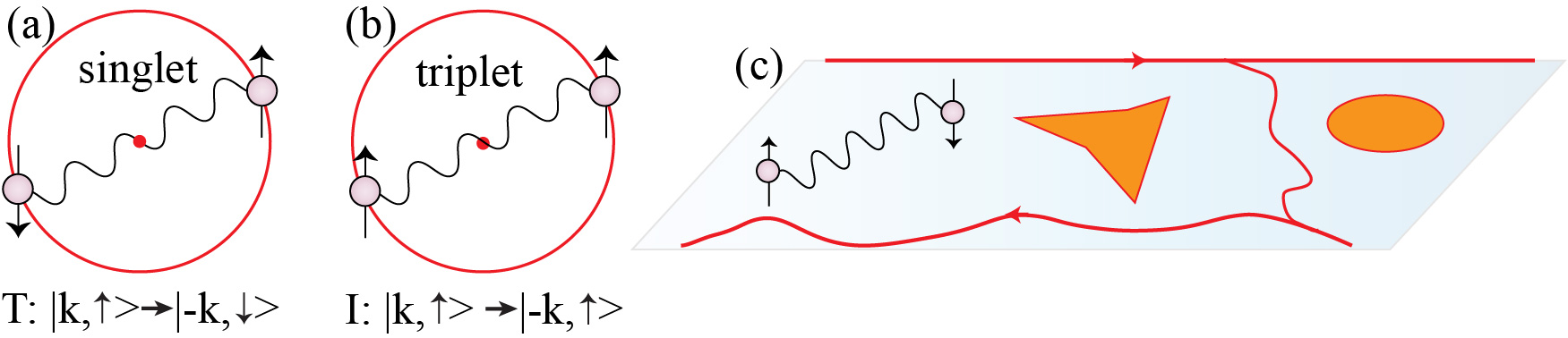}\caption{(a) The time-reversal operator $T$ maps a $|k,\uparrow>$ state to $|-k,\downarrow>$ state, which is the essential symmetry for spin singlet Cooper pairs. (b) The inversion operator $I$ maps a $|k,\uparrow>$ state to $|-k,\uparrow>$ state, which is the essential symmetry for spin triplet Cooper pairs.  (c) The edge states, CDW domain walls and other crystal grain boundaries etc. are not gapped out by the SC pairings.
			\label{fig1}}
	\end{center}
	\vskip-0.5cm
\end{figure*}

 In this paper, we suggest that because of the existence of TRSB,  the gapless topological edge states  are forbidden by symmetry from opening a gap by pairing in proximity to a conventional pairing.  Thus, those gapless states on the domains of the CDW or hidden orders remain gapless in the superconducting states.  Specifically, there are two key discrete symmetries in SCs to guarantee the presence of Cooper pairing, time-reversal $T$ and inversion symmetry $I$ \cite{sigrist,anderson1,anderson2}.  Since the $T$ maps a $|k,\uparrow>$ state to $|-k,\downarrow>$ state, the system at least contains time-reversal symmetry for the even-parity spin-singlet pairing formed by $(c_{k,\uparrow}c_{-k,\downarrow}-c_{k,\downarrow}c_{-k,\uparrow})$, as illustrated in Fig.\ref{fig1}(a). In the same spirit, the odd-parity spin-triplet pairing needs inversion symmetry $I$ owing to the fact that $I$ maps a $|k,\uparrow>$ state to $|-k,\uparrow>$ state, as illustrated in Fig.\ref{fig1}(b). These two symmetry conditions are known as the Anderson theorem \cite{sigrist,anderson1,anderson2}. 
 For AV$_3$Sb$_5$ SC cases,  due to TRSB, the normal states before the SC transition  breaks the $T$ symmetry. Therefore, the edge modes on CDW domain walls which break  the $T$ symmetry or other crystal grain boundaries  cannot be gapped out by the SC, as illustrated in Fig.\ref{fig1}(c). Hence, although the AV$_3$Sb$_5$ SC can be a conventional $s$-wave, it still contains gapless excitations. To demonstrate this physics, we construct a four band  tight-binding (TB) model, which faithfully captures the band structures of AV$_3$Sb$_5$ near Fermi energy. Since the inversion symmetry is always a good symmetry for  AV$_3$Sb$_5$ normal state from the recent second-harmonic generation measurement \cite{yuli}, the spin singlet pairing and spin triplet pairing should be separated. We will focus on the spin singlet in this work based on the decreasing Knight shift from NMR \cite{zhengli}.

\begin{figure}[htb]
	\centerline{\includegraphics[width=0.5\textwidth]{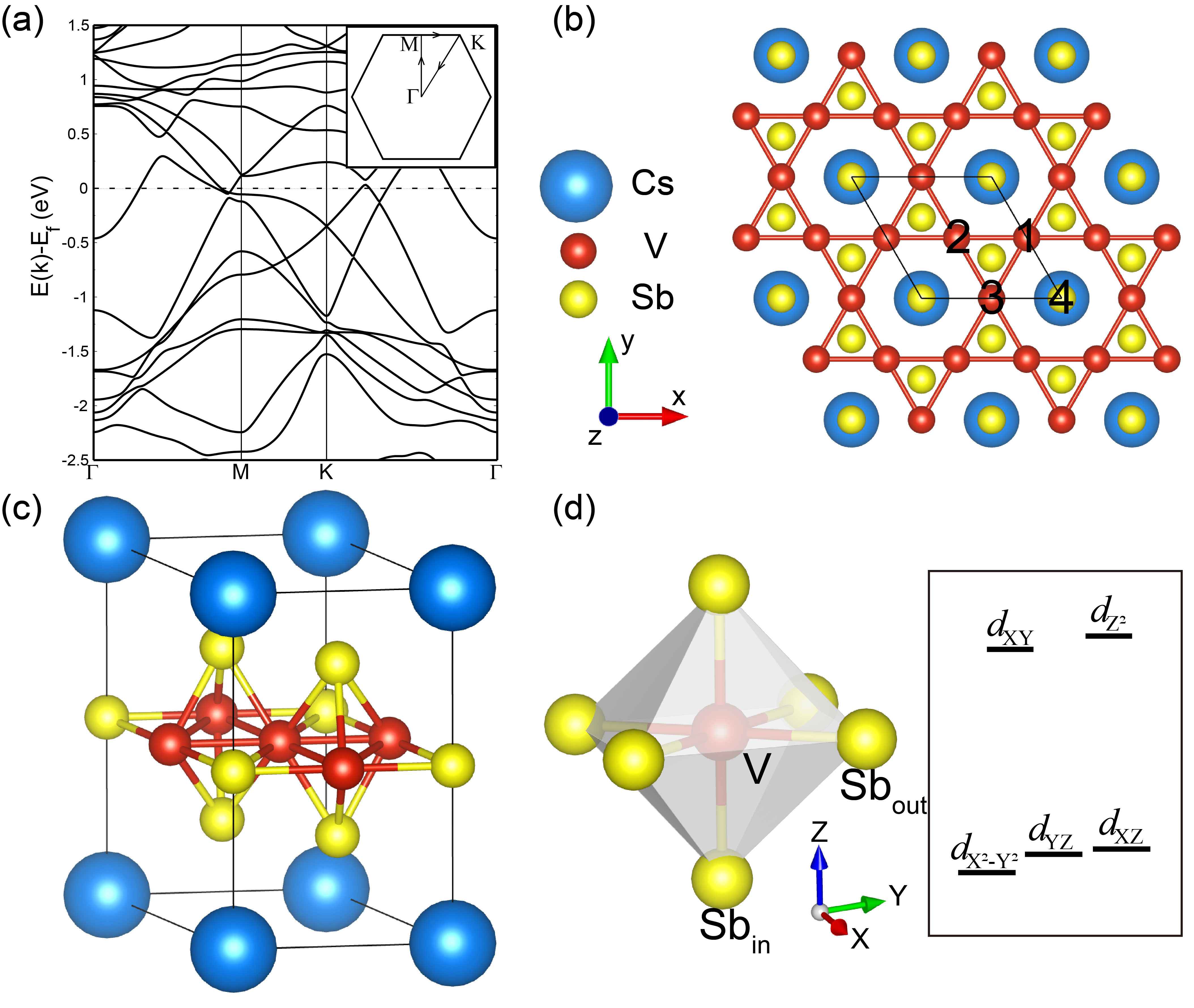}} \caption{(a) Band structure of CsV$_3$Sb$_5$ from DFT calculation with SOC. The inset shows the high symmetry $\textbf{k}$-path we use. (b) Crystal structure of CsV$_3$Sb$_5$ from the top view. The Wannier centers of Wannier functions in the TB model are labeled as 1-4. (c) Crystal structure of CsV$_3$Sb$_5$ from the another angle view. (d)
		The illustration of the VSb$_6$ octahedra complex under the local coordinate ($X-Y-Z$) and the schematic illustration for the crystal field splitting under this local coordinate. There are two types of Sb atoms: two Sb atoms are in the same plane with V atoms while the other four Sb atoms are out-of-plane. 
		\label{fig2} }
\end{figure}

Since all AV$_3$Sb$_5$ materials  have very similar band structures, we take CsV$_3$Sb$_5$ as an example.  From the density functional theory (DFT) calculations and ARPES measurements \cite{ortiz20,zhouxj}, there are multi-bands crossing the Fermi level, as shown in Fig.\ref{fig2}(a).   The crystal structure of  CsV$_3$Sb$_5$  is shown in Fig.\ref{fig2}(b-c). The DFT and ARPES results show that AV$_3$Sb$_5$ is a quasi-two-dimensional
metal, whose electronic physics is dominated by electrons from the V-Sb plane \cite{ortiz19,ortiz20,zhouxj}. In this  V-Sb plane, three V atoms  form a  kagome lattice and an additional Sb atom forms a triangle lattice locating at the V hexagonal center. Above and below this V-Sb plane, out-of-plane Sb atoms form two honeycomb lattices respectively.  Cs atoms form another  triangle lattice above or below these Sb honeycomb planes.

We will show that a four-band TB model can faithfully capture the main physics behind AV$_3$Sb$_5$ based on Wannierization and symmetry analysis. To understand the band structure, we first focus on the local atomic structure of V atoms. There are three V atoms in the kagome lattice's unit cell (labeled as 1-3, as indicated in Fig.\ref{fig2}(b)) and here we choose the local coordinate ($X-Y-Z$) for each site, as shown in Fig.\ref{fig2}(d). Let's take V-3 as an example to explain the definition of the local coordinate: as the global coordinate changes to the local coordinate, the axis $x/y/z$ turns to $Z/X/Y$ on V-3. By using $C_3$ symmetric operation, we can get the other two local coordinates on V-1 and V-2. In the local symmetric coordinate, the $Z$-axis always points to the in-plane Sb atom and the $Y$-axis is same as the $z$-axis in the global $x-y-z$ coordinate. The V atom can be considered as being coordinated in a distorted octahedron, whose point group symmetry is $D_{2h}$. The $D_{2h}$ crystal field leads to no degeneracies of all the five $d$-orbitals (Fig.\ref{fig2}(d)), consistent with our Wannierization result\cite{sm}. In the local coordinate, the energy of $d_{XY}$/$d_{Z^2}$ orbital is higher than that of $d_{X^2-Y^2}$/$d_{XZ}$/$d_{YZ}$ orbital, similar to the familiar $e_g$-$t_{2g}$ relationship in non-distorted octahedral crystal field.  

\begin{figure*}[ht]
	\centerline{\includegraphics[width=0.9\textwidth]{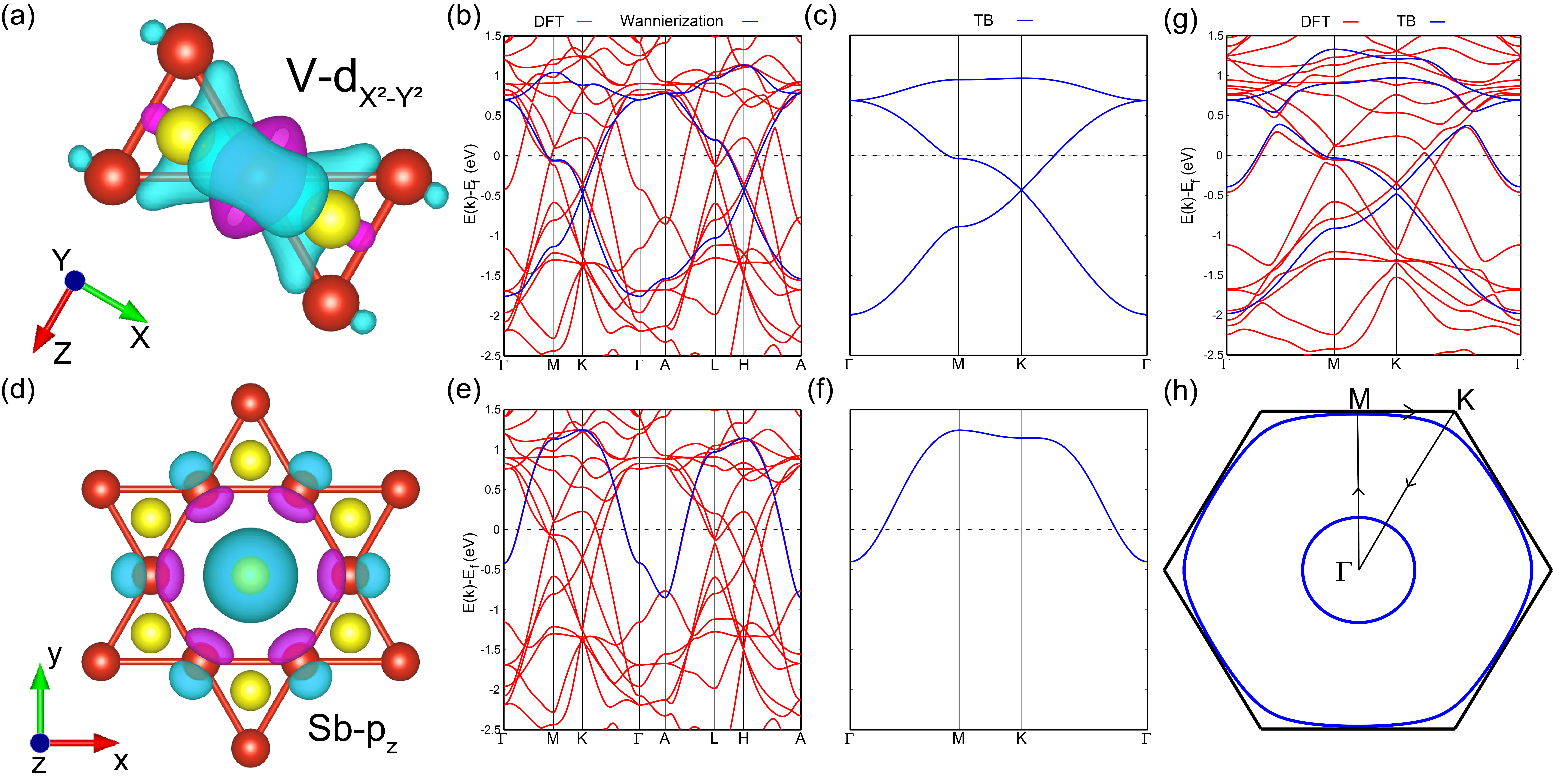}} \caption{(a) The $d_{X^2-Y^2}$-like Wannier function on V atom under the local $X-Y-Z$ coordinate. The other two $d_{X^2-Y^2}$-like Wannier functions are symmetric. (b) Band structures of CsV$_3$Sb$_5$ from DFT and Wannierization of three $d_{X^2-Y^2}$-like Wannier functions without SOC. (c) Band structure of the TB model with three $d_{X^2-Y^2}$-like Wannier functions. (d) The $p_z$-like Wannier function on in-plane Sb atom under the global $x-y-z$ coordinate. (e) Band structures of CsV$_3$Sb$_5$ from DFT and Wannierization of  the $p_z$-like Wannier function without SOC. (f) Band structure of the TB model with the $p_z$-like Wannier function. (g) Band structure of the TB model with SOC. (h) The Fermi surface of the TB model. 
		\label{fig3} }
\end{figure*}

From the DFT calculation and Wannierization, we find that the eigenstate of the von Hove(vH)  point is mainly from the local $d_{X^2-Y^2}$ orbital discussed above. More importantly, the $d_{X^2-Y^2}$ orbital of this vH is dominated by the single V sublattice. This feature is the reason why a single orbital model on kagome lattice is a reasonable starting model for these materials \cite{jxli,qhwang13,thomale13,feng2}. Namely, a minimal TB model based on the local $d_{X^2-Y^2}$ orbitals can capture the main physics of AV$_3$Sb$_5$, especially the vH around the Fermi level. Besides the V $d$-orbital, there is one additional electron pocket around $\Gamma$ point, which is  attributed to  the in-plane Sb's $p_z$ orbital. 
Without the spin-orbital coupling (SOC), there is no overlap between the in-plane Sb $p_z$ orbital and V $d_{X^2-Y^2}$-orbital. Hence, the $p_z$ band can be isolated from V $d_{X^2-Y^2}$ bands. 

Based on the above observation, we apply the maximally localized Wannier functions (MLWFs) method to get the minimal model for CsV$_3$Sb$_5$. Using the local V-centered $d_{X^2-Y^2}$-like Wannier function, a three band model on kagome lattice is obtained as shown in Fig.\ref{fig3}(b). The corresponding Wannier function is also plotted in Fig.\ref{fig3}(a). By comparing with the DFT results in Fig.\ref{fig3}(b), we find this three-band model well describes the main feature of vH points around the Fermi level and the Dirac cone at the K points. Additionally, using the in-plane-Sb-centered $p_z$-like Wanner function, a single $p_z$ band on triangle lattice is also obtained, as plotted in Fig.\ref{fig3}(e). Its Wannier function is also shown in Fig.\ref{fig3}(d). As shown in Fig.\ref{fig3}(e), the Wannierized $p_z$ band exactly agrees with the DFT calculation. Hence,  the $p_z$ band is isolated from other bands without SOC as discussed above. The SOC coupling terms can be further added in the minimal model using point group symmetry \cite{sm}.

The effective TB model in the basis of three symmetric $d_{X^2-Y^2}$-like Wannier functions (labeled as 1-3) and one $p_z$-like Wannier function (labeled as 4) to describe the in-plane electronic physics, as shown in Fig.\ref{fig3}(g).   Including the SOC, the TB Hamiltonian can be written as
\begin{eqnarray}
	H_{TB}=H_{V}+H_{Sb}+H_{SOC}.
\end{eqnarray}
Details of this Hamiltonian with SOC can be found in the supplemental material \cite{sm}.

\begin{figure}
	\begin{center}
		\fig{3.4in}{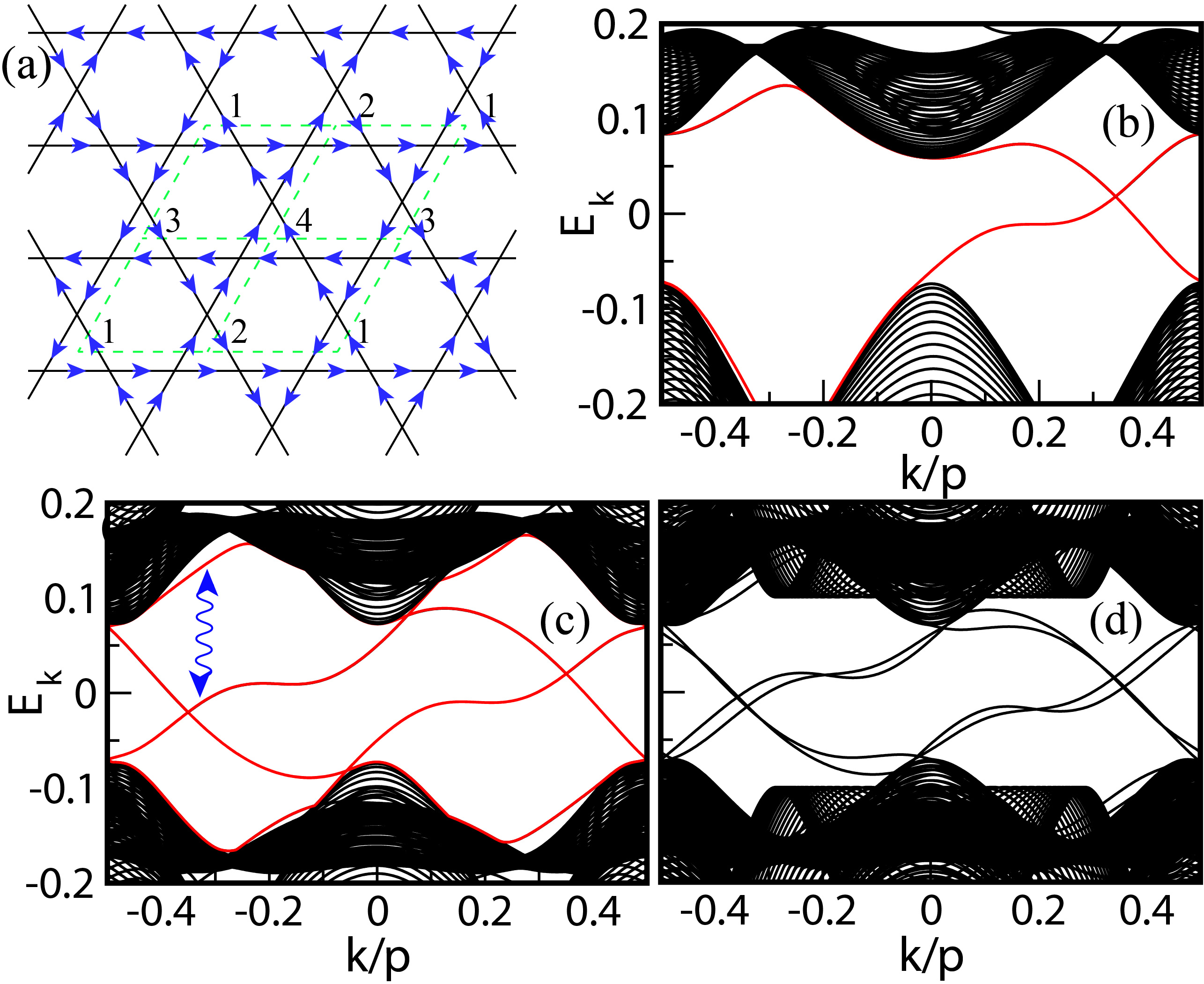}\caption{(a)  The hopping flux configuration for the chiral flux phase.  (b) The edge states spectrum for the chiral flux phase without Sb bands. (c) The BdG spectrum for the pairing state without Sb bands.  The energy difference $\delta_E$ between $E_{k,\uparrow}$ and $E_{-k,\downarrow}$ is on the order of $\Delta_{CFP}$. (d) The BdG spectrum for the pairing state  with Sb bands, where the SOC splits the in-gap bands. Here we take the CFP order parameter $\Delta_{CFP}$=0.2 eV and the onsite pairing for both V and Sb atoms to be $\Delta_{SC}$=0.1 eV and the chemical potential $\mu$=-0.1 eV. For the edge state calculation, we consider a system with translation invariance along $\textbf{a}_1$ direction and open boundary condition along $\textbf{a}_2$ direction with length of 100 lattice constant.
			\label{fig4}}
	\end{center}
	\vskip-0.5cm
\end{figure}

Motivated by the chiral CDW found by  magnetic-field dependent STM measurements \cite{topocdw} and its concurrence with AHE \cite{xhchen}, several TRSB flux states have been proposed to explain the phenomena\cite{feng,ronny,balents,feng2,nandkishore}. Among them, the lowest energy chiral flux phase (CFP) as shown in Fig.\ref{fig4}(a)\cite{feng,feng2} has gained support from recent $\mu$SR measurements \cite{yuli,musr1}.   In the single orbital model,  the CFP state has $C=2$  Chern number \cite{feng}.  The topological aspects of the CFP state in the minimal model remains the same as before.   To confirm this, we carry out an open boundary  calculation for the minimal model coupling with the CFP order. Note that we have ignored the Sb bands to avoid the complication.  As shown in Fig.\ref{fig4}(b), there are chiral edge states from non-zero Chern number inside the bulk gap. It is important to point out that due to TRSB and inversion symmetry breaking at the boundary, the edge state spectrum has no relation between $\textbf{k}$ and $-\textbf{k}$ anymore.

Now, we consider  superconductivity in this system.  The standard Bogoliubov-de Gennes (BdG) Hamiltonian can be written  as   
\begin{equation}
	H_{BdG}=\left(
	\begin{array}{cccc}
		H_{TB}(k)-\mu  & \mathbf{\hat{\Delta}(\mathbf{r})} \\
		\mathbf{\hat{\Delta}^\dagger(\mathbf{r})} & -H_{TB}^T(-k)+\mu
	\end{array}
	\right), \label{h}
\end{equation}
in the basis $\Psi_k=\left(c_{k\uparrow},c_{k\downarrow},c_{-k\uparrow}^{\dagger},c_{-k\downarrow}^{\dagger}\right)^T$, where $\mu$ is the chemical potential and $\mathbf{\hat{\Delta}(\mathbf{r})}$ is the  pairing function. We consider a standard $s$-wave function,  $\hat{\mathbf{\Delta}}(\mathbf{r})=\mathbf{\Delta(\mathbf{r})}is_y$ with $s_y$ the corresponding Pauli matrix in the spin space.
Since our symmetry analysis and discussion does not depend on details of the microscopic model, we take the on-site $s$-wave pairing as an example, namely $\mathbf{\Delta(\mathbf{r})}=\Delta_{SC}$. As shown in Fig.\ref{fig4}(c), the edge states remain gapless after the SC order is introduced. From Fig.\ref{fig4}(b) and Fig.\ref{fig4}(c), we can find that the energy difference $\delta E$ between $E_{k,\uparrow}$ of the $|k,\uparrow>$ state  and $E_{-k,\downarrow}$ of the $|-k,\downarrow>$ state is on the order of  the TRSB gap $\Delta_{CFP}$. Therefore, if the pairing gap $\Delta_{SC}$ is much smaller than $\delta E$, the  edge states are always gapless. This result is still valid  if we include the Sb band and the SOC term in the above calculation, as shown in Fig.\ref{fig4}(d).

Experimentally,  $\Delta_{CFP}$ is much larger than the $\Delta_{SC}$ in this family of materials. The $\mu$SR measurements show that the TRSB starts around $T \approx 70$K for CsV$_3$Sb$_5$ with the CDW transition temperature $T \approx 90$K \cite{yuli}. The SC transition temperature is much smaller with $T_c \approx 2.5$K \cite{ortiz20}.  Hence, regardless of the microscopic pairing form, we expect that there are always gapless excitations  in the edges, domain walls and other places of AV$_3$Sb$_5$ SC, where the TRSB order play a dominated role.  This result can explain the residual thermal conductivity, where the gapless excitations contribute the thermal conductivity like the conventional electrons. It is clear that the above result can be extended to the spin-triplet pairing bulk state.  In this case, the topological gapless states at the inversion symmetry broken structures can remain gapless.




In summary, we conclude that the AV$_3$Sb$_5$5 ground state contains gapless excitations owing to its TRSB normal state even if the bulk  superconductivity is fully gapped. To confirm this, a four-band TB model is constructed to capture electronic band structures of the materials,  using the V-centered $d_{X^2-Y^2}$ -like Wannier functions and the in-plane Sb-centered $p_z$-like Wannier function based on the technique of Wannierization and symmetry analysis.  In this case,  topological gapless excitations due to the TRSB  remain gapless when a fully gapped SC order parameter is induced.  Our proposal  can be justified or falsified easily by measuring the states on edges or domain boundaries.

We thank Shiyan Li, Li Yu and Jiaxin Yin for useful discussions. This work is supported by the Ministry of Science and Technology  (Grant
No. 2017YFA0303100), National Science Foundation of China (Grant No. NSFC-11888101), and the Strategic Priority Research Program of Chinese Academy of Sciences (Grant
No. XDB28000000). K.J. acknowledges support from the start-up grant of IOP-CAS.

\clearpage
\newpage

\begin{center}
    \Large
    Supplemental Material for "Gapless excitations inside the fully gapped kagome superconductors AV$_3$Sb$_5$"
\end{center}

\renewcommand\thefigure{S\arabic{figure}} 
\setcounter{figure}{0}
\renewcommand\thetable{S\arabic{table}} 
\setcounter{table}{0}
\renewcommand\theequation{S\arabic{equation}}
\setcounter{equation}{0}

\section{Computational details}
Our density functional theory calculations employ the Vienna ab initio simulation package (VASP) code\cite{kresse1996} with the projector augmented wave (PAW) method\cite{Joubert1999}. The Perdew-Burke-Ernzerhof (PBE)\cite{perdew1996} exchange-correlation functional is used in our calculations. The kinetic energy cutoff is set to be 600 eV for the expanding the wave functions into a plane-wave basis in VASP calcuations while the energy convergence criterion is $10^{-6}$ eV. The experimental crystal structures of \ce{AV3Sb5} (A=K,Rb,Cs) are adopted\cite{ortiz2019new}. The $\Gamma$-centered \textbf{k}-mesh is $12\times12\times7$.

We employ Wannier90\cite{mostofi2008wannier90,Marzari2012} to calculate maximally localized Wannier functions. We get our TB model's parameters by fitting the band structure from the Wannierization.

\section{Wannierization with V's $d$-orbitals and S\MakeLowercase{b}'s $p$-orbitals}
\begin{figure}[htb]
	\centerline{\includegraphics[width=0.5\textwidth]{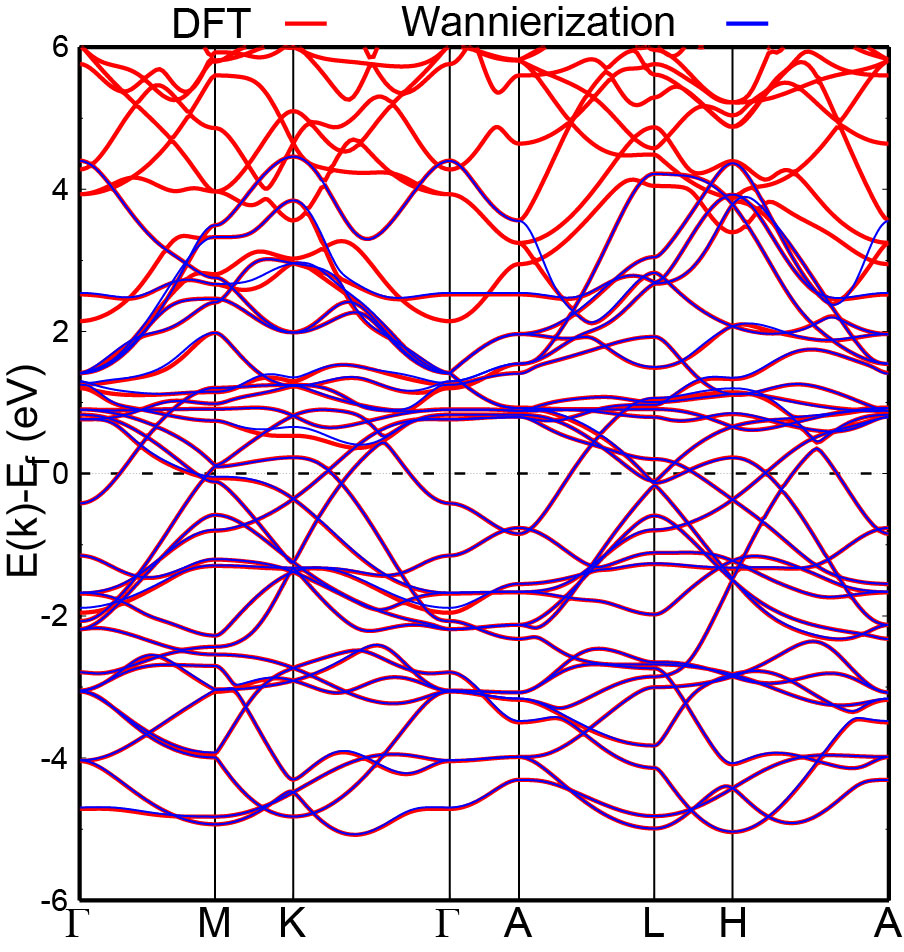}} \caption{(color online) Comparison between band structures from DFT and Wannierization with V's $d$-orbitals and Sb's $p$-orbitals in \ce{CsV3Sb5}.
		\label{DFT-wan} }
\end{figure}

\begin{figure*}[htb]
	\centerline{\includegraphics[width=0.9\textwidth]{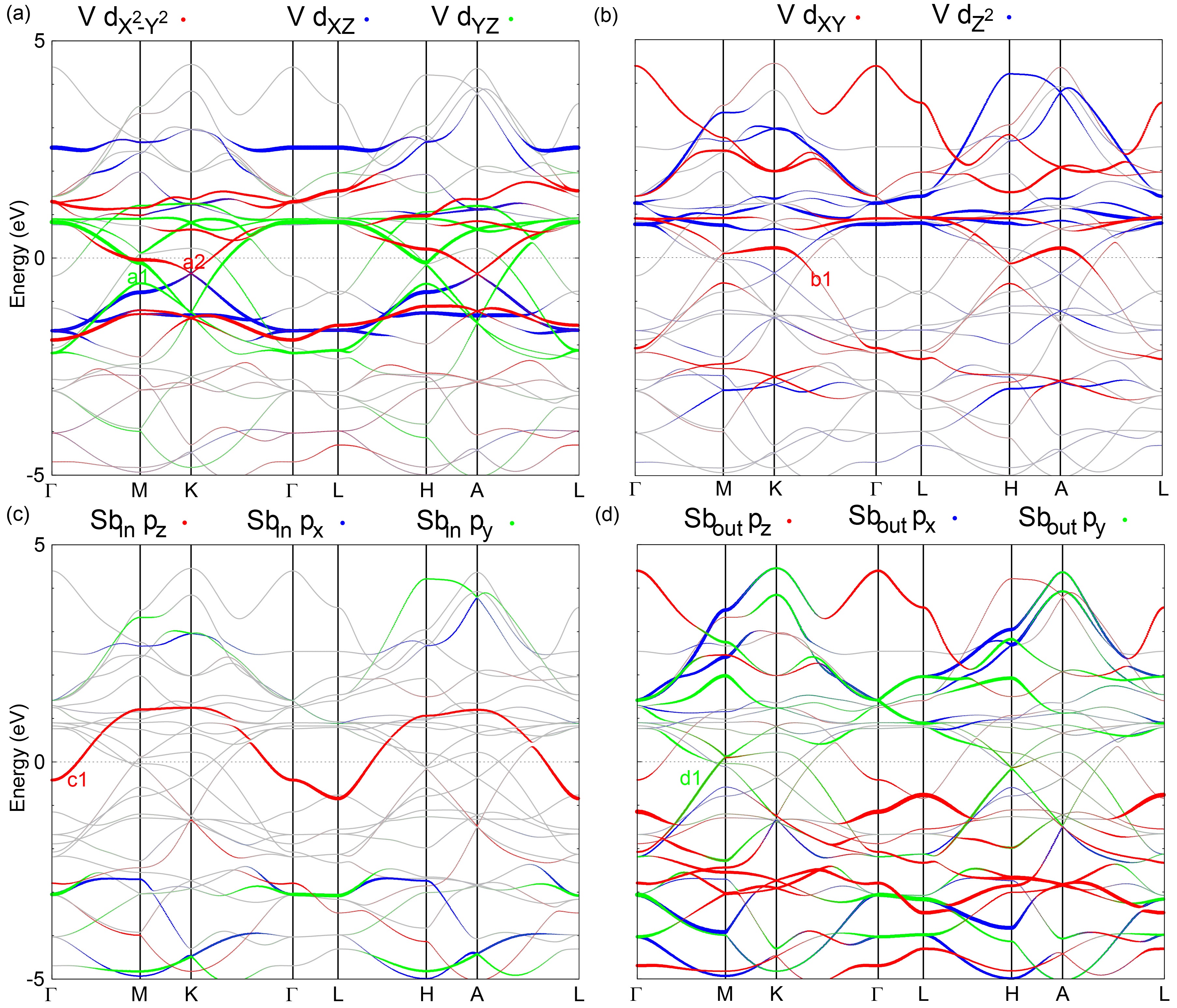}} \caption{(color online) \ce{CsV3Sb5}'s projected band structures without spin-orbital coupling (SOC) from Wannierization of V's $d$-orbitals and Sb's $p$-orbitals. The bands crossing the Fermi level are labeled as a1, a2, b1, c1 and d1, which are mainly composed by V-$d_{YZ}$ orbital, V-$d_{X^2-Y^2}$ orbital,  V-$d_{XY}$ orbital, Sb$_{in}$-$p_z$ orbital and Sb$_{out}$-$p_y$ orbital, respectively.
		\label{bands} }
\end{figure*}

Our Wannierization with V's $d$-orbitals and Sb's $p$-orbitals can perfectly reproduce band structure from DFT, as shown in FIG.\ref{DFT-wan}. In our local coordinate, there are five different parts which cross the Fermi level, as labeled in FIG.\ref{bands}. The a1/a2 part, forming a vH singularity at $M$ point and a Dirac cone at $K$ point, is mainly attributed to V-$d_{YZ/X^2-Y^2}$ orbital, as shown in FIG.\ref{bands}(a). The a1/a2 part crosses the Fermi level along $\Gamma-M$ line and $K-\Gamma$, respectively. The b1 band (FIG.\ref{bands}(b)), mainly composed of V-$d_{XY}$ orbital and also hybridizing with Sb$_{out}$-$p_y$ orbital, crosses the Fermi level along the $K-\Gamma$ line. Moreover, the c1 band in FIG.\ref{bands}(c) clearly indicates that Sb$_{in}$-$p_z$ orbital can be isolated around the Fermi energy and forms an electron pocket around the $\Gamma$ point. In FIG.\ref{bands}(d), the d1 band, chiefly contributed by Sb$_{out}$-$p_y$ orbital, crosses the Fermi level along $\Gamma-M$ line. It is worthy noting that our $\Gamma-M$ line is same as the direction of the global $y$-axis (FIG.\ref{4bands}(a)), thus Sb's $p_y$ component will change to a symmetric linear combination of Sb's $p_x$ and $p_y$ orbitals as rotate $M$ to $M'$ or $M''$. 

\clearpage
\section{Effective 4-band TB model}
\begin{figure}[h]
	\centerline{\includegraphics[width=0.5\textwidth]{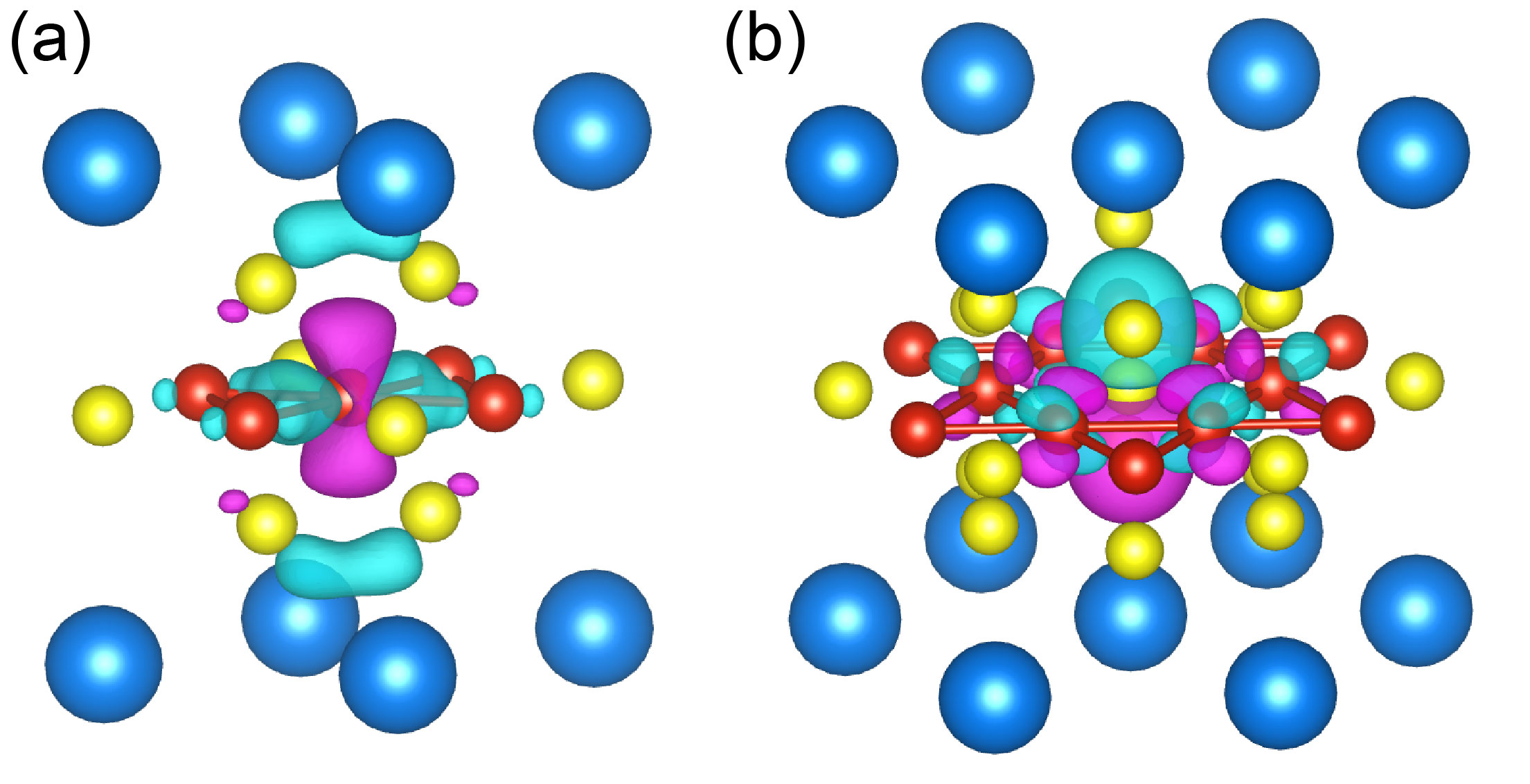}} \caption{(color online) (a) V-centered $d_{X^2-Y^2}$-like Wannier function and (b) Sb-centered $p_z$-like Wannier function in our 4-band TB model from another angle of view.
		\label{WF_PLOT} }
\end{figure}

We construct the effective TB model in the basis of three symmetric $d_{X^2-Y^2}$-like Wannier functions (labeled as 1-3) and one $p_z$-like Wannier function (labeled as 4) to describe the in-plane electronic physics:
\begin{equation}
\Psi^\dagger=(\Phi^\dagger_{d_{X^2-Y^2},V1},\Phi^\dagger_{d_{X^2-Y^2},V2},\Phi^\dagger_{d_{X^2-Y^2},V3},\Phi^\dagger_{p_{z},Sb_{in}})
\end{equation}
We plot these Wannier functions from another angle of view in FIG.\ref{WF_PLOT}.
The TB model can be written as a $4\times4$ Hermitian matrix ($H_{TB}$):
\begin{widetext}
\begin{eqnarray}\label{HTB}
H_{11}&=&\varepsilon_V+2t^1_{TNN}cos(-k_x/2+\sqrt{3}k_y/2)+2t^2_{TNN}(cos(k_x)+cos(k_x/2+\sqrt{3}k_y/2)),\nonumber \\
H_{22}&=&\varepsilon_V+2t^1_{TNN}cos(k_x/2+\sqrt{3}k_y/2)+2t^2_{TNN}(cos(k_x)+cos(-k_x/2+\sqrt{3}k_y/2)),\nonumber \\
H_{33}&=&\varepsilon_V+2t^1_{TNN}cos(k_x)+2t^2_{TNN}(cos(k_x/2+\sqrt{3}k_y/2)+cos(-k_x/2+\sqrt{3}k_y/2)),\nonumber \\
H_{44}&=&\varepsilon_{Sb}+2t^{Sb}_{NN}(cos(k_x)+cos(-k_x/2+\sqrt{3}k_y/2)+cos(k_x/2+\sqrt{3}k_y/2)) \nonumber \\
&+&2t^{Sb}_{SNN}(cos(3k_x/2-\sqrt{3}k_y/2)+cos(3k_x/2+\sqrt{3}k_y/2)+cos(\sqrt{3}k_y))\nonumber \\
H_{12}&=&2t_{NN}cos(k_x/2)+2t_{SNN}cos(\sqrt{3}k_y/2),\nonumber \\
H_{13}&=&2t_{NN}cos(k_x/4+\sqrt{3}k_y/4)+2t_{SNN}cos(3k_x/4-\sqrt{3}k_y/4),\nonumber \\
H_{23}&=&2t_{NN}cos(-k_x/4+\sqrt{3}k_y/4)+2t_{SNN}cos(3k_x/4+\sqrt{3}k_y/4),\nonumber \\
H_{14}&=&H_{24}=H_{34}=0.\nonumber\\
\end{eqnarray}
\end{widetext}

\begin{figure}[h]
	\centerline{\includegraphics[width=0.5\textwidth]{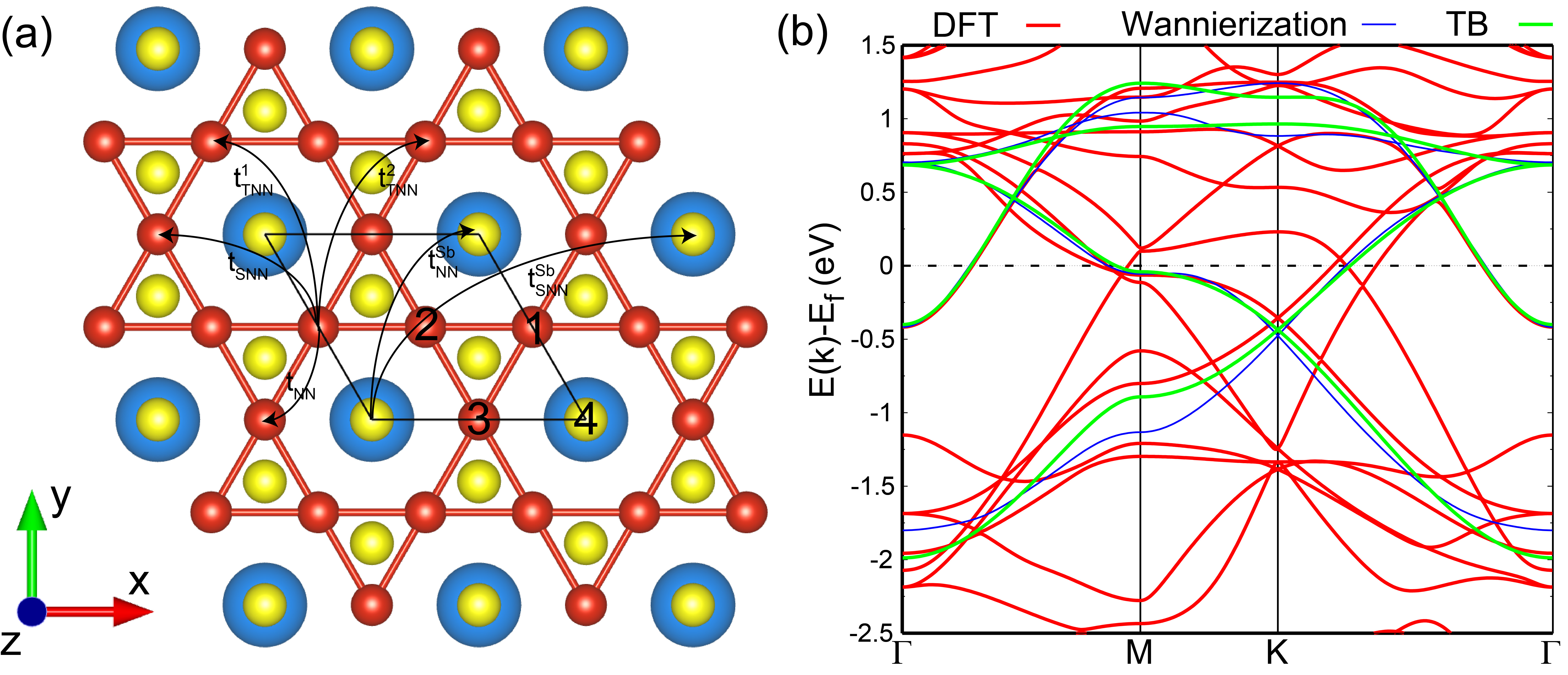}} \caption{(color online) (a) Definitions of hopping parameters in our TB model. (b) band structures of \ce{CsV3Sb5} from DFT, Wannierization and TB model without SOC.
		\label{4bands} }
\end{figure}

Our hopping parameters are truncated between third nearest neighbours (TNN) on V sites and second nearest neighbours on Sb sites, whose definitions are shown in FIG.\ref{4bands}(a). We get hopping parameters and on-site energies by fitting the band structure of Wannierization as in Ref.\cite{gu2021bacus2} (TABLE.\ref{tb_table}). 

\begin{table}[htb]
\caption{\label{tb_table}%
The parameters in our 4-band TB model. The unit is eV here.}
\begin{ruledtabular}
\begin{tabular}{cc}
 $\varepsilon_V$  &     -0.0480  \\
 $t_{NN}$ & -0.453 \\
 $t_{SNN}$& 0.00713 \\
 $t^1_{TNN}$ & -0.0377 \\
 $t^2_{TNN}$ & -0.0205 \\
 $\varepsilon_{Sb}$ & 0.831 \\
 $t^{Sb}_{NN}$ & -0.172 \\
 $t^{Sb}_{SNN}$& -0.0335 \\
\end{tabular}
\end{ruledtabular}
\end{table}

\clearpage
\section{spin-orbital coupling}
\begin{figure}[htb]
	\centerline{\includegraphics[width=0.5\textwidth]{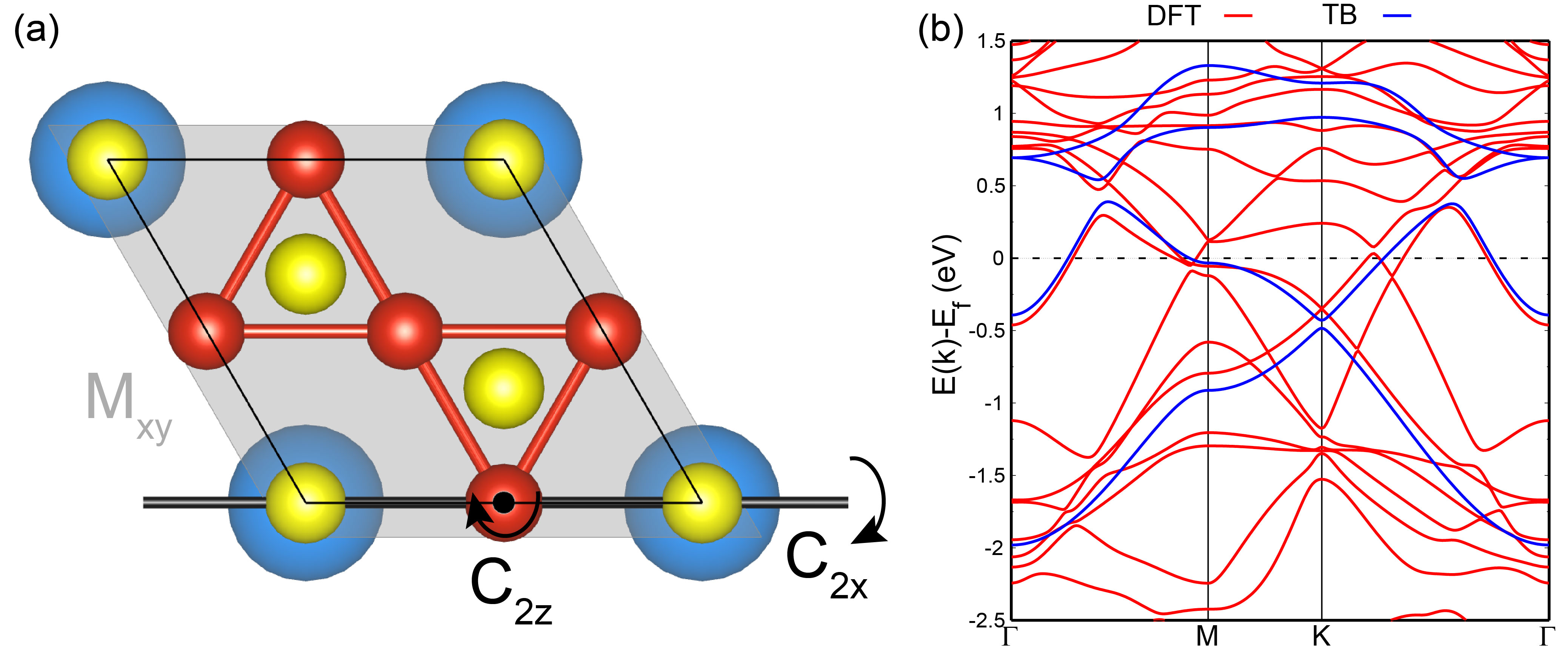}} \caption{(color online) (a) Schematic illustration for mirror symmetry $M_{xy}$ and rotation symmetry $C_{2x}/C_{2z}$ in \ce{CsV3Sb5}. (b) Band structures of \ce{CsV3Sb5} from DFT and the TB model  with SOC.  
		\label{SOC} }
\end{figure}

From main text, we can see that the couplings between $p_z$ and $d$ bands are mostly from the spin-orbital coupling (SOC). For example, a non-negligible gap between V's band and Sb's band along $\Gamma-M$ line owing to SOC, as shown in FIG.\ref{SOC}(b).  Based on point group symmetry, one can generated all possible SOC coupling terms in our minimal model. The procedure is following. Taking  V-3's orbital and its nearest neighbouring in-plane Sb's orbital as an example, the general form of SOC is
\begin{equation}
    H^{V_3-Sb}_{SOC}=a c_{V_3 \uparrow}^{+}c_{Sb\uparrow}+b c_{V_3 \uparrow}^{+}c_{Sb\downarrow}+c c_{V_3 \downarrow}^{+}c_{Sb\uparrow}+d c_{V_3 \downarrow}^{+}c_{Sb\downarrow}+h.c.
\end{equation}
This term can by simplified with time-reversal symmetry, rotation symmetry $C_{2x}$ and mirror symmetry $M_{xy}$, which are shown in FIG.\ref{SOC}(a). With time-reversal symmetry, we can get $a=d^*, b=-c*$, so the SOC term can be expressed as:
\begin{equation}
H^{V_3-Sb}_{SOC}=a c_{V \uparrow}^{+}c_{Sb\uparrow}+b c_{V \uparrow}^{+}c_{Sb\downarrow}-b^{*} c_{V \downarrow}^{+}c_{Sb\uparrow}+a^{*} c_{V \downarrow}^{+}c_{Sb\downarrow}+h.c..
\end{equation}
Then we can simplify the SOC term further by considering symmetry operators $C_{2x}$ and $M_{xy}$, as shown in FIG.\ref{SOC}(a).
The rotation symmetry $C_{2x}$ leads $a=-a^*=i\lambda, b=b*$ ($\lambda$ and $b$ are real numbers). Namely, the SOC term can be written as:
\begin{equation}
    H^{V_3-Sb}_{SOC}=i\lambda (c_{V_3 \uparrow}^{+}c_{Sb\uparrow}-c_{V_3 \downarrow}^{+}c_{Sb\downarrow})+b(c_{V_3 \uparrow}^{+}c_{Sb\downarrow}-c_{V_3 \downarrow}^{+}c_{Sb\uparrow})+h.c..
\end{equation}
With considering the mirror symmetry $M_{xy}$ , we can have $\lambda=0$. As a result, the SOC term between V3's orbital and Sb's orbital now is simplified as:
\begin{equation}
H^{V_3-Sb}_{SOC}=b(c_{V_3 \uparrow}^{+}c_{Sb\downarrow}-c_{V_3 \downarrow}^{+}c_{Sb\uparrow})+h.c..
\end{equation}
The rotation symmetry $C_{2z}$ leads the SOC term changes sign when switching two in-plane Sb atoms, so the SOC term on lattice becomes:
\begin{eqnarray}
&H&_{SOC}^{V_3-Sb}=\sum_{ij}b(c_{V_3 ij \uparrow}^{+}c_{Sb ij+\frac{\vec{a_{1}}}{2}\downarrow}-c_{V_3 ij \downarrow}^{+}c_{Sb ij+\frac{\vec{a_{1}}}{2} \uparrow}) \nonumber \\ 
&-&b(c_{V_3 ij \uparrow}^{+}c_{Sb ij-\frac{\vec{a_{1}}}{2}\downarrow}-c_{V_3 ij \downarrow}^{+}c_{Sb ij-\frac{\vec{a_{1}}}{2} \uparrow})+h.c..
\end{eqnarray}
With Fourier transformation, it becomes:
\begin{equation}
    H^{V_3-Sb}_{SOC}(\vec{k})=\sum_{\vec{k}}2i b sin(\frac{k_x}{2})(c_{V_3 \vec{k} \uparrow}^{+}c_{Sb \vec{k} \downarrow}-c_{V_3 \vec{k} \downarrow}^{+}c_{Sb \vec{k} \uparrow})+h.c.. \\
\end{equation}

By rotation symmetry, we can derive the remaining SOC term between V1/V2's orbital and Sb's orbital:
\begin{eqnarray}\label{V1V2_SOC}
H^{V_1-Sb}_{SOC}&=&\sum_{\vec{k}}2i b sin(\frac{-k_x+\sqrt{3}k_y}{4})((-\frac{1}{2}+i\frac{\sqrt{3}}{2})c_{V_1 \vec{k} \uparrow}^{+}c_{Sb \vec{k} \downarrow} \nonumber \\
&+&(\frac{1}{2}+i\frac{\sqrt{3}}{2})c_{V_1 \vec{k} \downarrow}^{+}c_{Sb \vec{k} \uparrow})+h.c, \\
H^{V_2-Sb}_{SOC}&=&\sum_{\vec{k}}2i b sin(\frac{k_x+\sqrt{3}k_y}{4})((\frac{1}{2}+i\frac{\sqrt{3}}{2})c_{V_2 \vec{k} \uparrow}^{+}c_{Sb \vec{k} \downarrow} \nonumber \\
&-&(\frac{1}{2}-i\frac{\sqrt{3}}{2})c_{V_2 \vec{k} \downarrow}^{+}c_{Sb \vec{k} \uparrow})+h.c.. \end{eqnarray}
With considering SOC , the full Hamiltonian is an $8\times8$ Hermitian matrix owing to the spin degree of freedom:
\begin{equation}
H=\begin{pmatrix}
H_{TB} & H_{SOC} \\
H^\dag_{SOC} & H_{TB}
\end{pmatrix}
\end{equation}
The matrix elements for SOC is in off-diagonal block ($H_{SOC}$):

\begin{eqnarray}\label{HSOC}
H_{18}&=&H_{4,5}=bsin(\frac{-k_x+\sqrt{3}k_y}{4})(-\sqrt{3}-i),\nonumber \\
H_{28}&=&H_{4,6}=bsin(\frac{k_x+\sqrt{3}k_y}{4})(-\sqrt{3}+i),\nonumber \\
H_{38}&=&H_{4,7}=bsin(\frac{k_x}{2})i, \nonumber \\
H_{15}&=&H_{16}=H_{17}=H_{25}=H_{26}=H_{27} \nonumber \\
&=&H_{35}=H_{36}=H_{37}=H_{48}=0.
\end{eqnarray}
Now there is only one parameter $b$. Here we estimate $b=0.1$ eV, and the corresponding band structure is plotted in FIG.\ref{SOC}(b). Gaps open because bands become the same double group representation.

\clearpage
\section{an alternative 7-band TB model}
\begin{figure}[htb]
	\centerline{\includegraphics[width=0.5\textwidth]{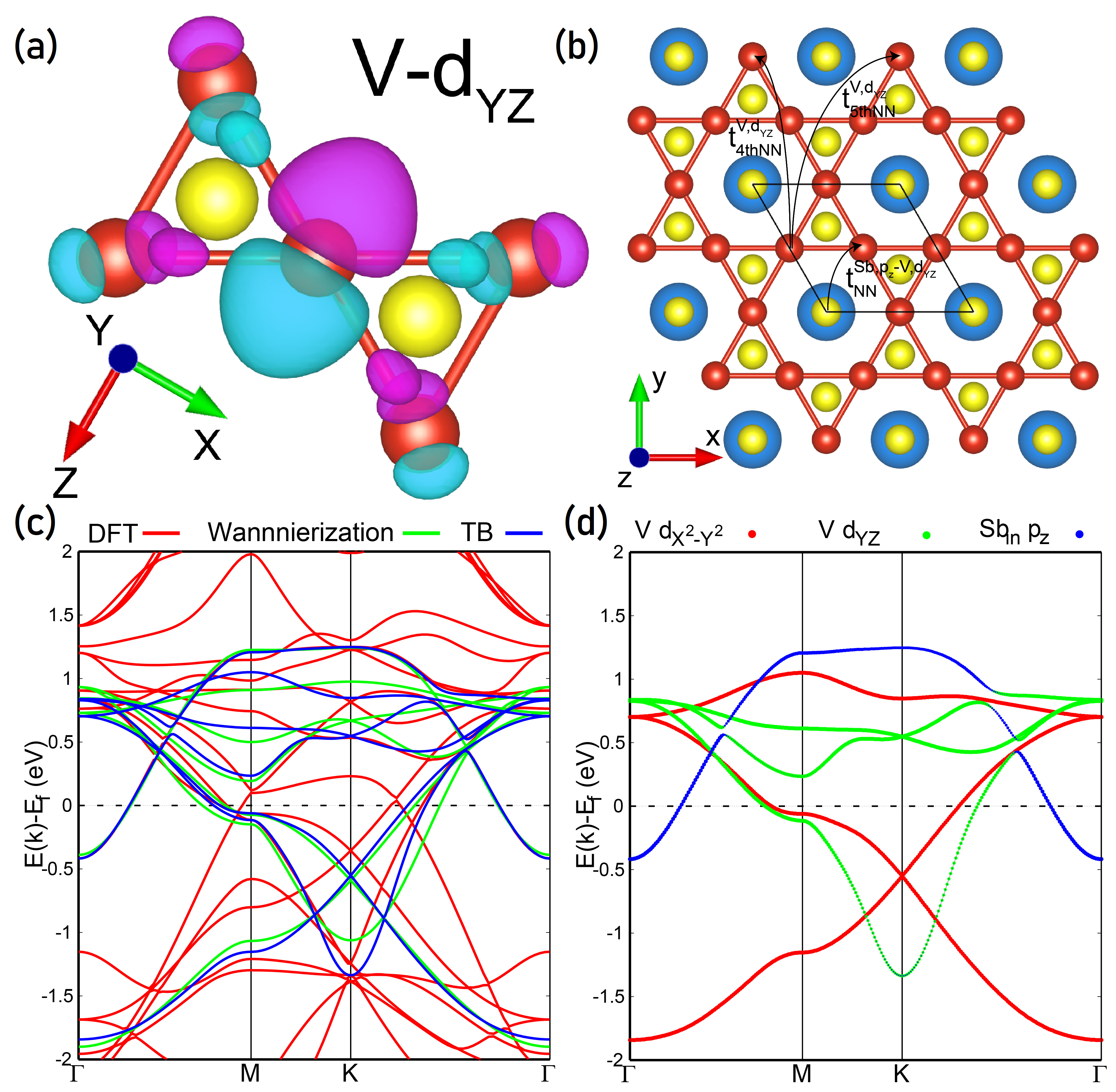}} \caption{(color online) (a) The $d_{YZ}$-like Wannier function on V atom under the local X-Y-Z coordinate. (b) Definitions of new hopping parameters in our 7-band TB model. (c) Band structures from DFT, Wannierization and 7-band TB model. (d) Fat band analysis from Wannierization and the sizes of the dots represent the weight of the projection.
		\label{7band} }
\end{figure}
As mentioned before (FIG.\ref{bands}), there are two vH points near the Fermi level, which indicates a two-orbital model\cite{kang2021twofold,xxwu2021pairing}. In our local coordinate, the lower vH point is mainly composed by V's $d_{YZ}$ orbital. Therefore, we can alternatively construct a 7-band TB model which can describe both in the basis:
\begin{widetext}
\begin{equation}
\Psi^\dagger=(\Phi^\dagger_{d_{X^2-Y^2},V1},\Phi^\dagger_{d_{X^2-Y^2},V2},\Phi^\dagger_{d_{X^2-Y^2},V3},\Phi^\dagger_{p_{z},Sb_{in}},\Phi^\dagger_{d_{YZ},V1},\Phi^\dagger_{d_{YZ},V2},\Phi^\dagger_{d_{YZ},V3}).
\end{equation}
\end{widetext}
The V-centered $d_{YZ}$-like Wannier function is plotted in FIG.\ref{7band}(a). So we similarly get a faithful Wannierization result with 7 initial projectors, as shown in FIG.\ref{7band}(c). Owing to the symmetry, the V-centered $d_{X^2-Y^2}$-like Wannier function does not hybridize with the Sb$_{in}$-centered $p_z$-like Wannier function, while the V-centered $d_{YZ}$-like Wannier function does, as shown in our fat band analysis (FIG.\ref{7band}(d)).

The $7\times7$ TB model has the following structure:
\begin{equation}
    H^{7band}_{TB}=\begin{pmatrix}
    \text{\Large ${H^{d_{X^2-Y^2}}}$} & \begin{matrix} 0 \\ 0 \\ 0 \end{matrix} &  \begin{matrix} 0 & 0 & 0 \\ 0 & 0 & 0 \\ 0 & 0 & 0 \end{matrix} \\
    \begin{matrix} 0 & 0 & 0 \end{matrix} & H^{p_z} & V^{{p_z}-{d_{YZ}}} \\
    \begin{matrix} 0 & 0 & 0 \\ 0 & 0 & 0 \\ 0 & 0 & 0 \end{matrix} & V^{\dagger{p_z}-{d_{YZ}}} & \text{\Large ${H^{d_{YZ}}}$}
    \end{pmatrix}
\end{equation}
The analytical terms of ${H^{d_{X^2-Y^2}}}$ and $H^{p_z}$ in 7-band TB model are same as those in the 4-band TB model. To get a decent fitting result, we need to consider longer hopping between V's 4/5th nearest neighbouring sites, as shown in FIG.\ref{7band}(b). The adding terms in $H^{7band}_{TB}$ can be written as:
\begin{widetext}
\begin{eqnarray}\label{7band_TB}
H_{55}&=&\varepsilon^{d_{YZ}}_{V}+2t^{V,d_{YZ},1}_{TNN}cos(-k_x/2+\sqrt{3}k_y/2)+2t^{V,d_{YZ},2}_{TNN}(cos(k_x)+cos(k_x/2+\sqrt{3}k_y/2)),\nonumber \\
H_{66}&=&\varepsilon^{d_{YZ}}_{V}+2t^{V,d_{YZ},1}_{TNN}cos(k_x/2+\sqrt{3}k_y/2)+2t^{V,d_{YZ},2}_{TNN}(cos(k_x)+cos(-k_x/2+\sqrt{3}k_y/2)),\nonumber \\
H_{77}&=&\varepsilon^{d_{YZ}}_{V}+2t^{V,d_{YZ},1}_{TNN}cos(k_x)+2t^{V,d_{YZ},2}_{TNN}(cos(k_x/2+\sqrt{3}k_y/2)+cos(-k_x/2+\sqrt{3}k_y/2)),\nonumber \\
H_{56}&=&2t^{V,d_{YZ}}_{NN}cos(k_x/2)+2t^{V,d_{YZ}}_{SNN}cos(\sqrt{3}k_y/2)\nonumber \\
&+&2t^{V,d_{YZ}}_{4thNN}(cos(k_x+\sqrt{3}k_y/2)+cos(k_x-\sqrt{3}k_y/2))+2t^{V,d_{YZ}}_{5thNN}cos(3k_x/2),\nonumber \\
H_{57}&=&2t_{NN}^{V,d_{YZ}}cos(k_x/4+\sqrt{3}k_y/4)+2t^{V,d_{YZ}}_{SNN}cos(3k_x/4-\sqrt{3}k_y/4)\nonumber \\
&+&2t^{V,d_{YZ}}_{4thNN}(cos(5k_x/4+\sqrt{3}k_y/4)+cos(-k_x/4+3\sqrt{3}k_3/4))+2t^{V,d_{YZ}}_{5thNN}cos(3k_x/4+3\sqrt{3}k_y/4),\nonumber \\
H_{23}&=&2t^{V,d_{YZ}}_{NN}cos(-k_x/4+\sqrt{3}k_y/4)+2t^{V,d_{YZ}}_{SNN}cos(3k_x/4+\sqrt{3}k_y/4)\nonumber \\
&+&2t^{V,d_{YZ}}_{4thNN}(cos(k_x/4+3\sqrt{3}k_y/4)+cos(5k_x/4-\sqrt{3}k_y/4))+2t^{V,d_{YZ}}_{5thNN}cos(-3k_x/4+3\sqrt{3}k_y/4),\nonumber \\
H_{45}&=&-2it_{NN}^{Sb,p_z-V,d_{YZ}}sin(-k_x/4+\sqrt{3}k_y/4),\nonumber \\
H_{46}&=&2it_{NN}^{Sb,p_z-V,d_{YZ}}sin(k_x/4+\sqrt{3}k_y/4),\nonumber \\
H_{47}&=&-2it_{NN}^{Sb,p_z-V,d_{YZ}}sin(k_x/2).\nonumber\\
\end{eqnarray}
\end{widetext}
This 7-band TB model can faithfully fit the DFT-calculated band structure, as shown in FIF.\ref{7band}(c). This result also reveals the two orbital nature of the two vH singularites around the Fermi level: one vH point is composed by local $d_{X^2-Y^2}$-like Wannier function, whose NN hopping is dominating; another vH is composed by local $d_{YZ}$-like Wannier function, whose TNN hopping is the biggest.

\begin{table}[htb]
\caption{\label{7band-tb_table}%
The parameters in our 7-band TB model. The unit is eV here.}
\begin{ruledtabular}
\begin{tabular}{cc}
 $\varepsilon^{d_{X^2-Y^2}}_{V}$  &     -0.0946  \\
 $t^{V,d_{X^2-Y^2}}_{NN}$ & -0.4661 \\
 $t^{V,d_{X^2-Y^2}}_{SNN}$& 0.0278 \\
 $t^{V,d_{X^2-Y^2},1}_{TNN}$ & -0.008 \\
 $t^{V,d_{X^2-Y^2},2}_{TNN}$ & -0.0091 \\
 $\varepsilon^{p_z}_{Sb}$ & 0.8215 \\
 $t^{Sb,p_z}_{NN}$ & -0.1813 \\
 $t^{Sb,p_z}_{SNN}$& -0.0204 \\
 $\varepsilon^{d_{YZ}}_{V}$  &  0.3619  \\
 $t^{V,d_{YZ}}_{NN}$ & 0.1112 \\
 $t^{V,d_{YZ}}_{SNN}$& -0.1168 \\
 $t^{V,d_{YZ},1}_{TNN}$ & 0.0077 \\
 $t^{V,d_{YZ},2}_{TNN}$ & 0.1315 \\
 $t^{V,d_{YZ}}_{4thNN}$ & 0.0356 \\
 $t^{V,d_{YZ}}_{5thNN}$ & -0.0802 \\
 $t_{NN}^{Sb,p_z-V,d_{YZ}}$ & -0.0071 \\
\end{tabular}
\end{ruledtabular}
\end{table}

\begin{figure*}[ht]
	\centerline{\includegraphics[width=0.9\textwidth]{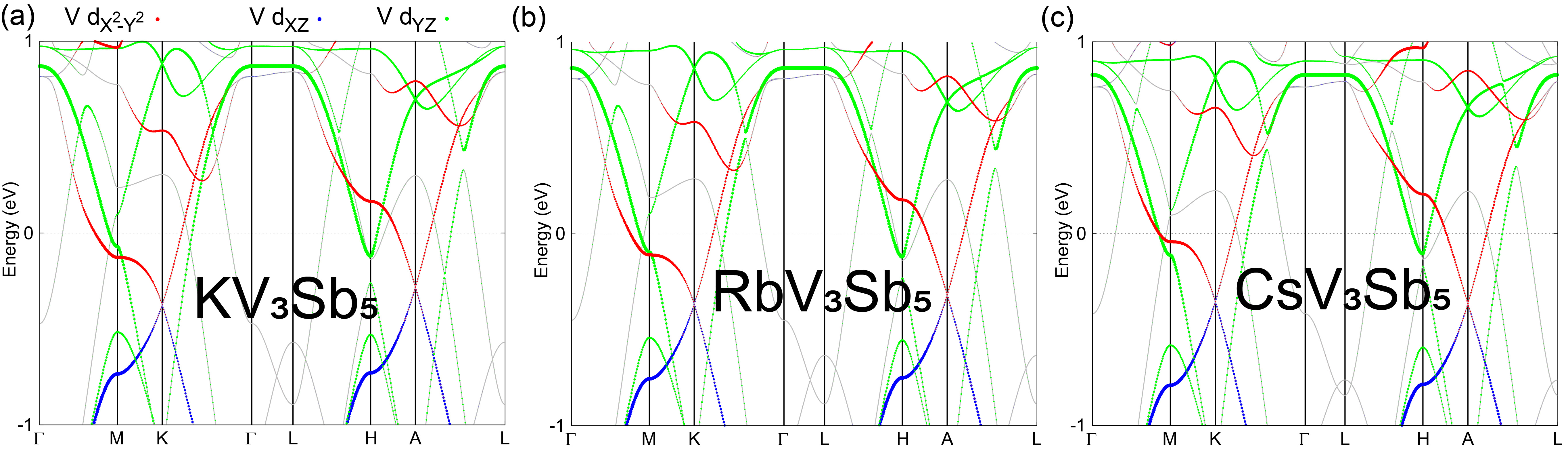}} \caption{(color online) Projected band structures without SOC from Wannierization of V's $d$-orbitals and Sb's $p$-orbitals for (a)\ce{KV3Sb5}, (b)\ce{RbV3Sb5} and (c)\ce{CsV3Sb5} . 
		\label{KVSbands} }
\end{figure*}

It is worthy noting that the $d_{YZ}$-composed vH will be the nearest vH to the Fermi level in \ce{KV3Sb5}/\ce{RbV3Sb5}, as show in FIG.\ref{KVSbands}. This might be a significant difference between \ce{CsV3Sb5} and \ce{KV3Sb5}/\ce{RbV3Sb5}.

%

\end{document}